# AndorEstimator: Android based Software Cost Estimation Application

Muhammad Zubair Asghar
Institute of Computing and Information Technology
Gomal University, D.I.Khan, Pakistan

Ammara Habib
Institute of Computing and Information Technology
Gomal University, D.I.Khan, Pakistan

Anam Habib
Institute of Computing and Information Technology
Gomal University, D.I.Khan, Pakistan

Syeda Rabail Zahra
Institute of Computing and Information Technology
Gomal University, D.I.Khan, Pakistan

Sadia Ismail
Institute of Computing and Information Technology
Gomal University, D.I.Khan, Pakistan

***Abstract—:*** **The main aim of the proposed system is to assist the software development team to estimate the cost, effort and maintenance of the project under development. Android-based platform, namely MIT App Inventor is used for the development of application, which contains visual block programming language. The current study has following uniqueness of (1)Accuracy of results,(2)user friendly environment(3)no such application is available on android platform to the best of our knowledge. Questionnaire regarding CoCoMo model is developed and circulated by using objective qualitative method.** ***Findings:*** **The estimation module of our application is quite important with respect to facilitating the students of software engineering for performing CoCoMo-based cost estimation easily, and enabling the software developers for performing software cost estimation easily. The cost estimator based on CoCoMo model is developed on android platform however, to the best of our knowledge no such application is available. This system can be used by business and educational stakeholders, such as students, software developers, and business organizations**

**Keywords—CoCoMo model; App Inventor; Cost estimation; Android**

## I. INTRODUCTION

Development of android-based Conversion and Estimation application can assist the stakeholders related to business and educational sector through the use of smart phones and tablets. The students and software engineers can estimate their project cost, effort, and person per month in the early development of software life cycle.

For the software project management, the software cost assessment is mandatory to reduce the risks and to better analyze the software development process. The accuracy in estimation of cost also help in decision making. So for this purpose CoCoMo model was developed by using genetic model and ant colony optimization approach to develop the software product by optimizing the current coefficients. In order to find the exact and accurate estimation genetic algorithm is widely used [1]. However in the current era of android based there is a need to develop an android based software cost estimation.

In most of the software effort estimation methodologies that include CoCoMo model were unsuccessful to provide a reliable reference for project manager due to its lack of accuracy.

So the Fuzzy expert – CoCoMo model was developed that provide following facilities such as vital information about the estimated effort and also has the ability to not only amalgamate the effort assessment and risk assessment activities into the initial planning phase [2].

To estimate the size, cost and schedule of software projects many refined methods and models are existing. Although for agile software projects the capability to flawlessly predict the software cost of web based software is still doubtful. So Agile MOW approach is presented here in this paper to evaluate effort and cost of software development using agile methodology that is developed for web based projects [3].

In the project planning, software effort estimation is one of the pre-eminent step to be carried out. In order to develop efficient and effective software's accurate estimates are required. For some decades many cost estimation methods have been provided by the software's researchers. As the COCOMO II model is the simplest model so it is commonly used model among different other models. Also there is no clear benchmark to design neural network model and the fuzzy approach is hard to use. So genetic algorithm has the ability to be a justifiable additional tool for software effort estimation. For optimizing the current coefficients of COCOMO II model this works aim to propose a genetic algorithm in order to achieve more accuracy in estimation [4].

In software development process accurate cost estimation of software project is one of the necessary accomplishment. So it help both the customer and the project manager to make reasonable decisions during project execution. On the other hand there is not much difference between Real Time Software System (RTSS) cost estimation and maintenance cost estimation but for RTSS some critical factors are considered like response time of software for input and processing time to give correct output [5].

For the competitiveness of the software companies the precision and reliability of the estimation of the estimation of the software product is very important. In the management of the software products, good estimates plays a very important rule. So a machine learning method is introduced in this paper which focus to compare machine learning techniques for software effort estimation and to show that robust confidence intervals for the effort estimation can be successfully built [6].

There is an offline application developed to perform CoCoMo based software cost estimation user makes input required for the CoCoMo I calculation and results are obtained [7]. Now a days the use of android is increasing rapidly so android based application for CoCoMo based software cost estimation is required.

The CoCoMo II based calculator is developed to perform cost estimation on the basis of inputs given by the user [8]. As this is an online calculator which require continuous internet connection so this deficiency can be overcome through Android application.

The application, available at [9] is used to perform the basic CoCoMo calculations in all modes: organic, semidetached, embedded.

Another web based system is developed for performing CoCoMo II calculations. The system is available at [10]. As the android application can easily be use for performing CoCoMo II calculations without a need to be online.

The existing literature [11-18] of different studies in text mining are conducted for the development of estimation applications The expansion of android-based application have changed the entire life style of individuals with maximum dependency on the hand-held devices both for entertainment as well as learning. Therefore development of android-based application for software cost estimation is required to facilitate users.

## II. Methodology

The proposed scheme consists of a main module of Software Estimation using CoCoMo model. It is further divided into sub components particularly: (1) CoCoMo I with basic model, (2) CoCoMo I with intermediate model, (3) CoCoMo I with detailed model, (4) CoCoMo II with early design model, (5) CoCoMo II with Post architecture model.

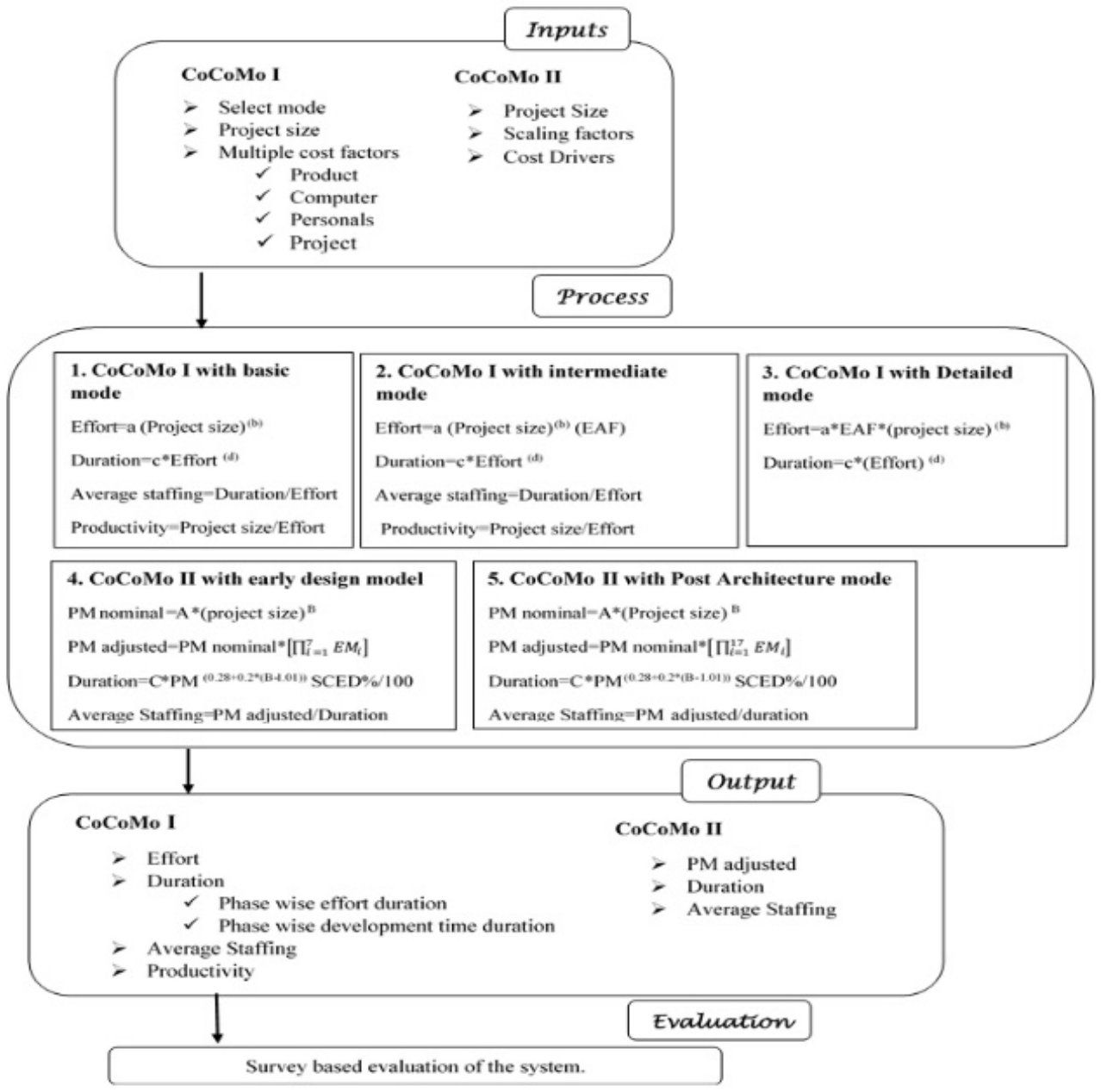

Fig. 1. The proposed system

Fig. 1 presents the execution of proposed system. In the first section inputs are given that are required by the users. The second section is the main section of proposed system in which computation is done where different formulas are used for processing. In the effort and duration equation of CoCoMo I model with basic model, intermediate model and detailed model the term a, b, c and d are constants whose values are given below:

TABLE I. Basic model

| Software Projects | Constant values | | | |
|---|---|---|---|---|
| | *a* | *b* | *c* | *d* |
| Organic | 2.4 | 1.05 | 2.5 | 0.38 |
| Semidetached | 3.0 | 1.12 | 2.5 | 0.35 |
| Embedded | 3.6 | 1.20 | 2.5 | 0.32 |

TABLE II. Intermediate model

| Software Projects | Constant values | | | |
|---|---|---|---|---|
| | *a* | *b* | *c* | *d* |
| Organic | 3.2 | 1.05 | 2.5 | 0.38 |
| Semidetached | 3.0 | 1.12 | 2.5 | 0.35 |
| Embedded | 2.8 | 1.20 | 2.5 | 0.32 |

TABLE III. Detailed model

| Software Projects | Constant values | | | |
|---|---|---|---|---|
| | *a* | *b* | *c* | *d* |
| Organic | 3.2 | 1.05 | 2.5 | 3.8 |
| Semidetached | 3.0 | 1.12 | 2.5 | 3.5 |
| Embedded | 2.8 | 1.20 | 2.5 | 3.2 |

Since in Fig. 1 EAF is the *Effort Adjustment Factor* obtained from the cost drivers. Similarly in CoCoMo II model the term A and C are also constants the values of whom are A=2.94 and C=3.67 where B is a scaling factor and its values is

$$B=0.91+0.1\Sigma^{7}_{i=1}SF_j \quad (1)$$

Where in (1) SF stands for *Scaling Factor*.

$$\Pi^{7}_{i=1}EM_i \quad (2)$$

Equation (2) shows that there are 7 cost drivers and

$$\Pi^{17}_{i=1}EM_i \quad (3)$$

Hence (3) shows that there are 17 cost drivers where EM means *Effort multiplier* .SCED is also a cost driver and it ranges from very low to very high. The third section represents the output of proposed system and survey based evolution is conducted in fourth section.

Now the flow chart of the proposed system is given below:

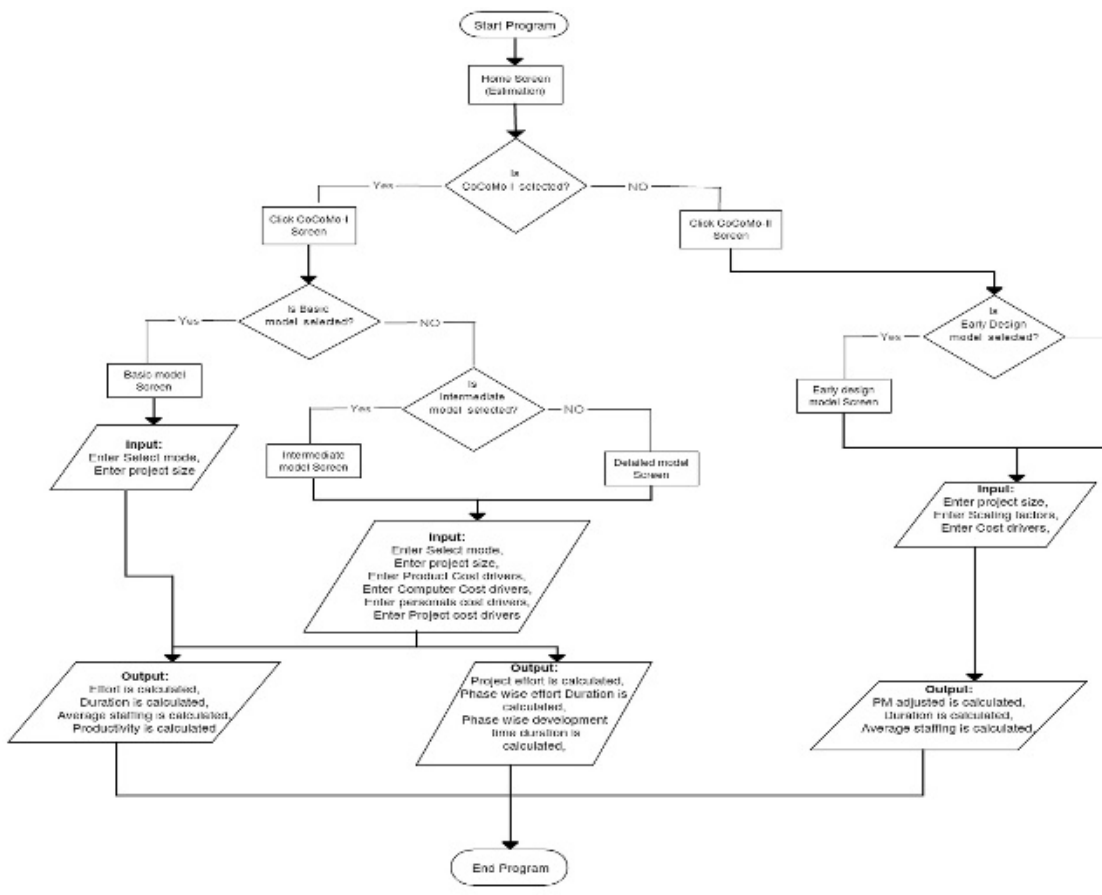

Fig. 2. Flowchart of proposed system

### A. CoCoMo I with basic model

The basic CoCoMo I model grants immediate and rough estimates of software cost on time. Users get the needed output of effort, duration, average staffing and productivity by entering the values of select mode and project size.

Algorithm 1. Computation of CoCoMo-I with Basic model

**Input:** Select mode, Project Size

**Output:** Effort, Duration, Average Size, Productivity

**Begin:**
1. If Select mode=Organic then
{
2. $Effort \leftarrow 2.4 \times (Project\ Size)^{1.05}$
3. $Duration \leftarrow 2.5 \times (effort)^{0.38}$
4. $Average\ Staffing \leftarrow {duration}/{effort}$
5. $Productivity \leftarrow {project\ size}/{effort}$
}
6. If Select mode=Semidetached then
{
7. $Effort \leftarrow 3 \times (Project\ Size)^{1.12}$
8. $Duration \leftarrow 2.5 \times (effort)^{0.35}$
9. $Average\ Staffing \leftarrow {duration}/{effort}$
10. $Productivity \leftarrow {project\ size}/{effort}$
}
11. If Select mode=Embedded then
{
12. $Effort \leftarrow 3.6 \times (Project\ Size)^{1.20}$
13. $Duration \leftarrow 2.5 \times (effort)^{0.32}$
14. $Average\ Staffing \leftarrow {duration}/{effort}$
15. $Productivity \leftarrow {project\ size}/{effort}$
}

**End**

### B. CoCoMo I with intermediate model

Intermediate model is an expansion of basic model because it enhances the features of basic model. Firstly, the user choose one of the selection mode (that is organic, semidetached and embedded) then he inputs the value of project size along with various cost drivers specially: (1) Product, (2) Computer, (3) Personals, and (4) Project. These inputs then find out effort, duration, average staffing and productivity which is the required output.

Algorithm 2. Computation of CoCoMo-I with Intermediate model

**Input:** Select mode, Project Size, Product, Computer, Personals, Project

**Output:** Effort, Duration, Average Size, Productivity

**Begin:**
1. If Select mode=Organic then
{
2. $Effort \leftarrow 3.2 \times (Project\ Size)^{1.05}\ (EAF)$
3. $Duration \leftarrow 2.5 \times (effort)^{0.38}$
4. $Average\ Staffing \leftarrow {duration}/{effort}$
5. $Productivity \leftarrow project\ size \div effort$
}
6. If Select mode=Semidetached then
{
7. $Effort \leftarrow 3.0 \times (Project\ Size)^{1.12}$
8. $Duration \leftarrow 2.5 \times (effort)^{0.35}$
9. $Average\ Staffing \leftarrow {duration}/{effort}$
10. $Productivity \leftarrow {project\ size}/{effort}$
}
11. If Select mode=Embedded then
{
12. $Effort \leftarrow 2.8 \times (Project\ Size)^{1.20}$
13. $Duration \leftarrow 2.5 \times (effort)^{0.32}$
14. $Average\ Staffing \leftarrow {duration}/{effort}$

15. $Productivity \leftarrow {}^{project\ size}/_{effort}$
}
**End**

*C. CoCoMo I with detailed model*

The detail model covers all the attributes of intermediate model, with an additional feature of calculating phase wise effort duration and phase wise development time duration of the required software. When users make selection from the selection mode (that is. organic, semidetached, and embedded), enters the value of projects size and different cost drivers namely: (1) Product, (2) Computer, (3) Personals, and (4) Project then effort, duration, phase wise effort distribution and phase wise development time duration are determined.

Algorithm 3. Computation of CoCoMo-I with Detailed model

**Input:** Select mode, Project Size

**Output:** Effort, Duration, Phase wise effort distribution, Phase wise development time duration.

**Begin:**
1. If Select mode=Organic then
{
2. $Effort \leftarrow 2.4 \times (Project\ Size)^{1.05}$
3. $Duration \leftarrow 2.5 \times (effort)^{0.38}$
}
4 If Select mode=Semidetached then
{
5. $Effort \leftarrow 3 \times (Project\ Size)^{1.12}$
6. $Duration \leftarrow 2.5 \times (effort)^{0.35}$
}
7. If Select mode=Embedded then
{
8. $Effort \leftarrow 3.6 \times (Project\ Size)^{1.20}$
9. $Duration \leftarrow 2.5 \times (effort)^{0.32}$
10. Call mode ();
}
11. Proc mode
{
12. If select mode=organic small
{
13. Display phase wise effort distribution
14. $Plan\ \&\ Requirment \leftarrow effort \times 0.06$
15. $System\ design \leftarrow effort \times 0.16$
16. $Detailed\ design \leftarrow effort \times 0.26$
17. $Module\ code\ \&\ Test \leftarrow effort \times 0.42$
18. $Integration\ \&\ Testing \leftarrow effort \times 0.16$
19. Display phase wise development time duration
20. $Plan\ \&\ Requirment \leftarrow Duration \times 0.1$
21. $System\ design \leftarrow Duration \times 0.19$
22. $Detailed\ design \leftarrow Duration \times 0.24$
23. $Module\ code\ \&\ Test \leftarrow Duration \times 0.39$
24. $Integration\ \&\ Testing \leftarrow Duration \times 0.18$
}
25. Else
{
26. Display phase wise effort distribution
27. $Plan\ \&\ Requirment \leftarrow effort \times 0.06$
28. $System\ design \leftarrow effort \times 0.16$
29. $Detailed\ design \leftarrow effort \times 0.24$
30. $Module\ code\ \&\ Test \leftarrow effort \times 0.38$
40. $Integration\ \&\ Testing \leftarrow effort \times 0.22$
41. Display phase wise development time duration
42. $Plan\ \&\ Requirment \leftarrow Duration \times 0.12$
43. $System\ design \leftarrow Duration \times 0.19$
44. $Detailed\ design \leftarrow Duration \times 0.21$
45. $Module\ code\ \&\ Test \leftarrow Duration \times 0.34$
46. $Integration\ \&\ Testing \leftarrow Duration \times 0.26$
}
47. If select mode=Semidetached medium
{
48. Display phase wise effort distribution
49. $Plan\ \&\ Requirment \leftarrow effort \times 0.07$
50. $System\ design \leftarrow effort \times 0.17$
51. $Detailed\ design \leftarrow effort \times 0.25$
52. $Module\ code\ \&\ Test \leftarrow effort \times 0.33$
53. $Integration\ \&\ Testing \leftarrow effort \times 0.25$
54. Display phase wise development time duration
55. $Plan\ \&\ Requirment \leftarrow Duration \times 0.2$
56. $System\ design \leftarrow Duration \times 0.26$
57. $Detailed\ design \leftarrow Duration \times 0.21$
58. $Module\ code\ \&\ Test \leftarrow Duration \times 0.27$
59. $Integration\ \&\ Testing \leftarrow Durtion \times 0.26$
}
60. Else
{
61. Display phase wise effort distribution
62. $Plan\ \&\ Requirment \leftarrow effort \times 0.07$
63. $System\ design \leftarrow effort \times 0.17$
64. $Detailed\ design \leftarrow effort \times 0.24$
65. $Module\ code\ \&\ Test \leftarrow effort \times 0.31$
66. $Integration\ \&\ Testing \leftarrow effort \times 0.28$
67. Display phase wise development time duration
68. $Plan\ \&\ Requirment \leftarrow Duration \times 0.22$
69. $System\ design \leftarrow Duration \times 0.27$
70. $Detailed\ design \leftarrow Duration \times 0.19$
71. $Module\ code\ \&\ Test \leftarrow Duration \times 0.25$
72. $Integration\ \&\ Testing \leftarrow Duration \times 0.29$
}
73. If select mode=Embedded large
{
74. Display phase wise effort distribution
75. $Plan\ \&\ Requirment \leftarrow effort \times 0.08$
76. $System\ design \leftarrow effort \times 0.18$
77. $Detailed\ design \leftarrow effort \times 0.25$
78. $Module\ code\ \&\ Test \leftarrow effort \times 0.26$
79. $Integration\ \&\ Testing \leftarrow effort \times 0.31$
80. Display phase wise development time duration
81. $Plan\ \&\ Requirment \leftarrow Duration \times 0.36$

82. $System\ design \leftarrow Duration \times 0.36$
83. $Detailed\ design \leftarrow Duration \times 0.18$
84. $Module\ code\ \&\ Test \leftarrow Duration \times 0.18$
85. $Integration\ \&\ Testing \leftarrow Duration \times 0.28$
}
86. Else
{
87. Display phase wise effort distribution
88. $Plan\ \&\ Requirment \leftarrow effort \times 0.08$
89. $System\ design \leftarrow effort \times 0.18$
90. $Detailed\ design \leftarrow effort \times 0.24$
91. $Module\ code\ \&\ Test \leftarrow effort \times 0.24$
92. $Integration\ \&\ Testing \leftarrow effort \times 0.34$
93. Display phase wise development time duration
94. $Plan\ \&\ Requirment \leftarrow Duration \times 0.4$
95. $System\ design \leftarrow Duration \times 0.38$
96. $Detailed\ design \leftarrow Duration \times 0.16$
97. $Module\ code\ \&\ Test \leftarrow Duration \times 0.16$
98. $Integration\ \&\ Testing \leftarrow Duration \times 0.3$
}
}
**End**

*D. CoCoMo II with early design model*

In the primary stages of software project there is slight information about the project size and its nature. So, early design model is used for basic estimates of project's cost and duration. When user inputs the value of (1) project size, (2) scaling factors and (3) multiple cost drivers then he/she gets the desired output of (1) PMnormal, (2) PM adjusted, (3) Duration and (4) Average staffing.

Algorithm 4**.** Calculation of CoCoMo-II with Early design model

**Input:** Project Size, Scaling factors, Cost Drivers

**Output:** PM adjusted, Duration, Average Staffing

**Begin:**
{
1. $PM\ nominal = a \times (Project\ Size)^B$
2. $PM\ ajusted = PM\ noinal \times [\pi_{i=1}^{7} EM_i]$
3. $Duration = 3.67 \times PM^{(0.28+0.2\times(B-1.01))}\dfrac{SCED\%}{100}$
4. $Average\ Saffing = {}^{PM\ ajusted}/_{Duration}$

}
**End**

*E. CoCoMo II with post architecture model*

Post architecture is the most comprehensive model of CoCoMo-II. When the final structure of project is developed then post architecture model is applied and it is also used for maintaining the software product. This model contain the inputs such as: (1) project size, (2) scaling factors and (3) multiple cost drivers which are then used to get the needed output of (1) PMnormal, (2) PM adjusted, (3) duration and (4) average staffing.

Algorithm 5**.** Calculation of CoCoMo-II with Post Architecture model

**Input:** Project Size, Scaling factors, Cost Drivers

**Output:** PM adjusted, Duration, Average Staffing

**Begin:**
{
1. $PM\ nominal = a \times (Project\ Size)^B$
2. $PM\ ajusted = PM\ noinal \times [\pi_{i=1}^{17} EM_i]$
3. $Duration = 3.67 \times PM^{(0.28+0.2\times(B-1.01))}\dfrac{SCED\%}{100}$
4. $Average\ Saffing = {}^{PM\ ajusted}/_{Duration}$

}
**End**

## III. Experimental Setup

The experimental setup section presents details about the implementation and evaluation of the proposed system. As described earlier, we developed the software using MIT App inventor and tested the apps in Blue stack emulator. To evaluate the effectiveness of proposed system, a web-based survey is conducted.

*A. Implementation*

In Fig. 3, the code block event handler is used in which two procedures are called. In procedure we use if then statement in which we use labels, textboxes and also a spinner code block is used, also some of the blocks are drop from the math Block editor such as Multiply and division.

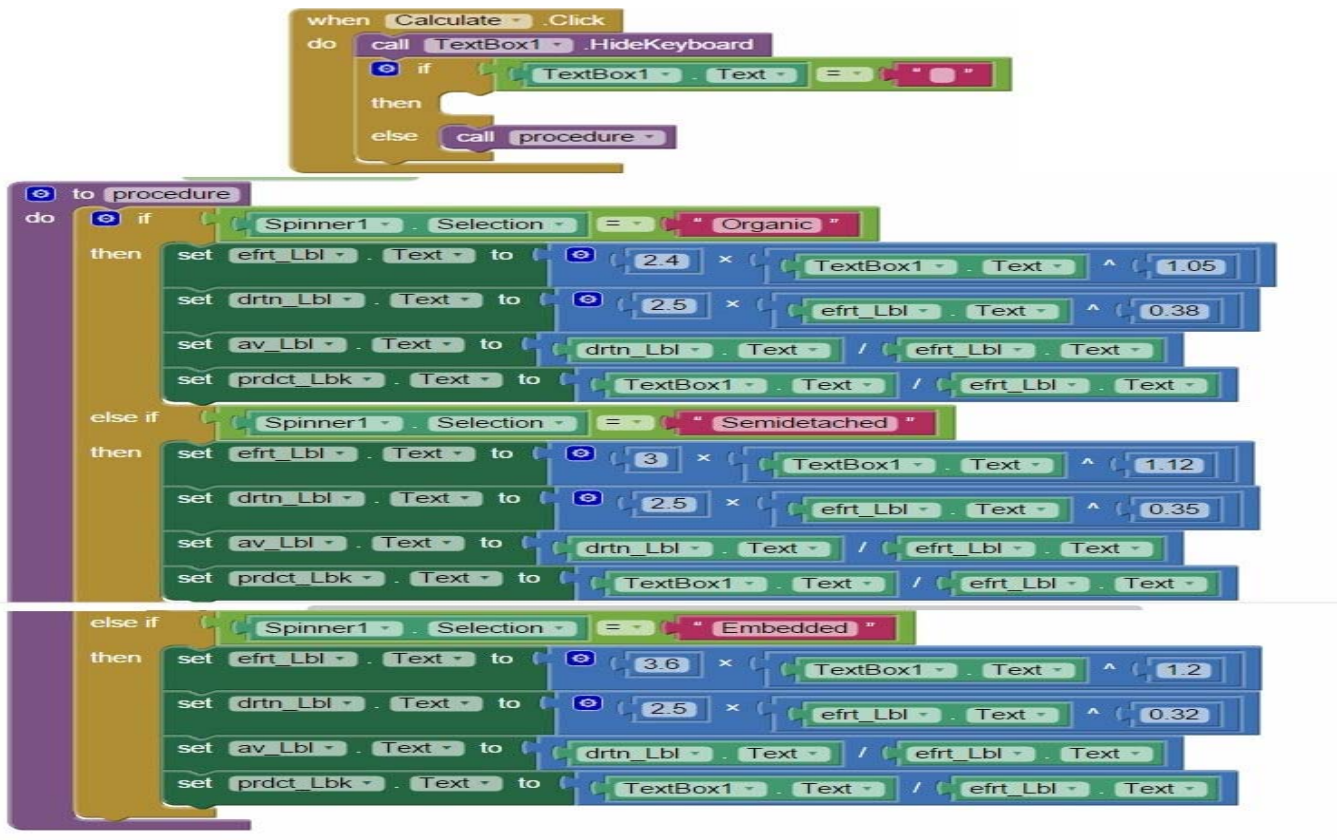

Fig. 3. Code block for CoCoMo I with basic model

Fig. 4 presents the intermediate code blocks which consists of if else statement, procedure, set of labels and also textboxes to execute the instructions.

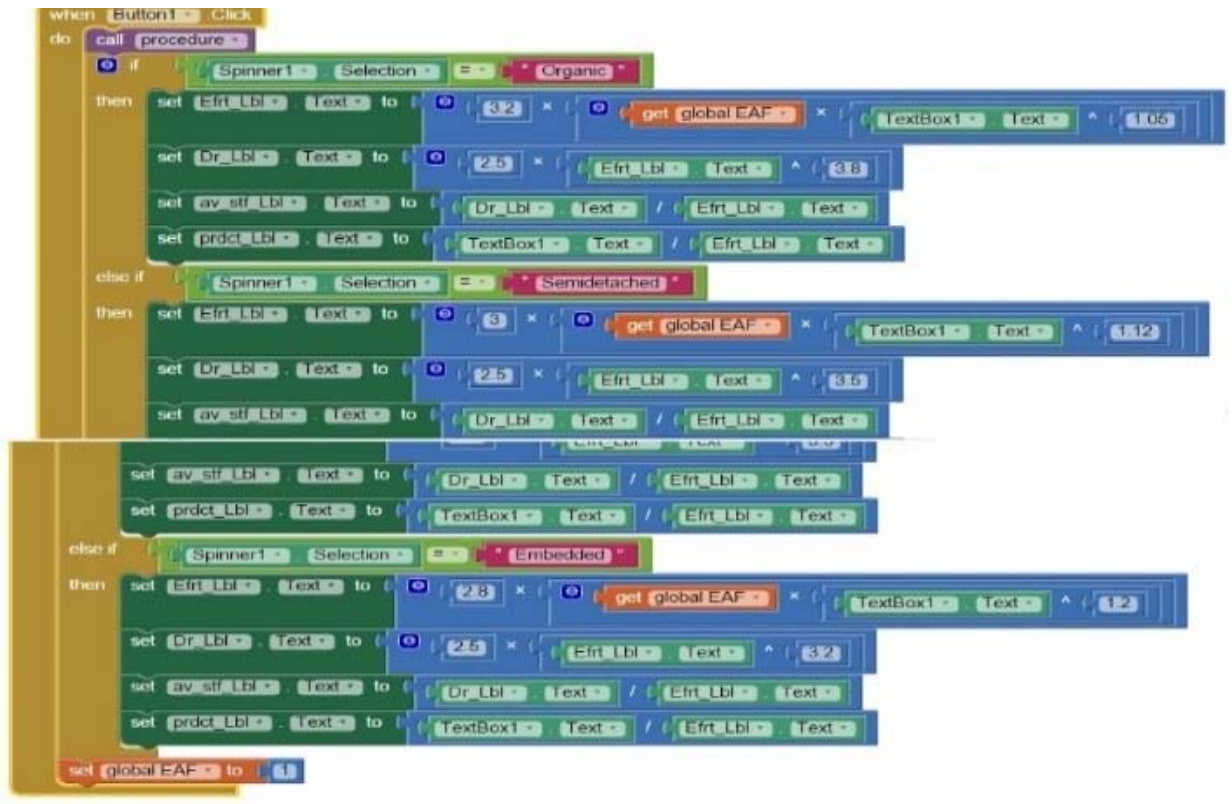

Fig. 4. Code block for CoCoMo I with intermediate model

In the detailed model, Button1 is an event handler which calls two procedures, it also contains labels textboxes and variables. The button contains a spinner component which gives a list of choices from which the user has to make a selection.

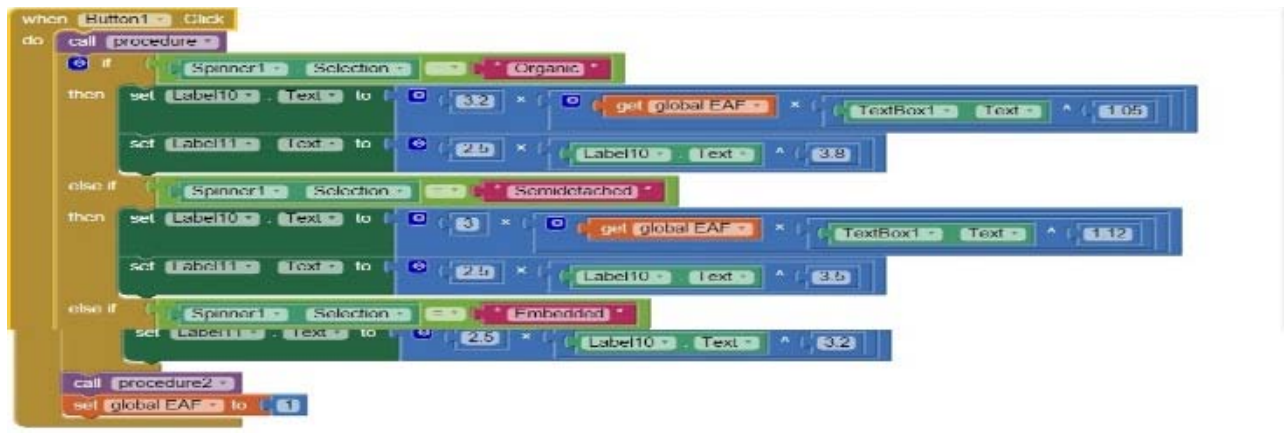

Fig. 5. Code block for CoCoMo I with detailed model

In Fig. 6 the code blocks for early design model that have two procedures, multiple variables and also textboxes are used.

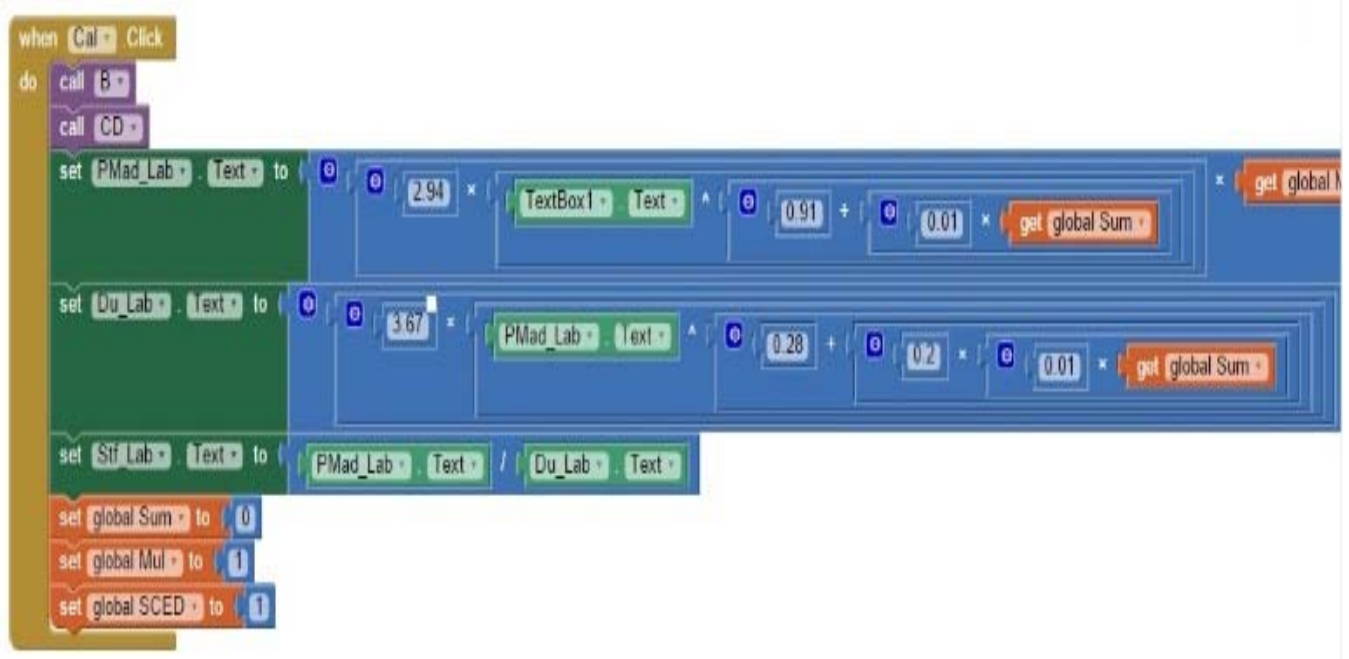

Fig. 6. Code block for CoCoMo II with early design model

When calculate button is clicked then a set of instruction inside the event handler are executed. All these instructions are executed in sequence. So it consists of two procedures, set of labels and variables. Fig. 7 shows the following code blocks.

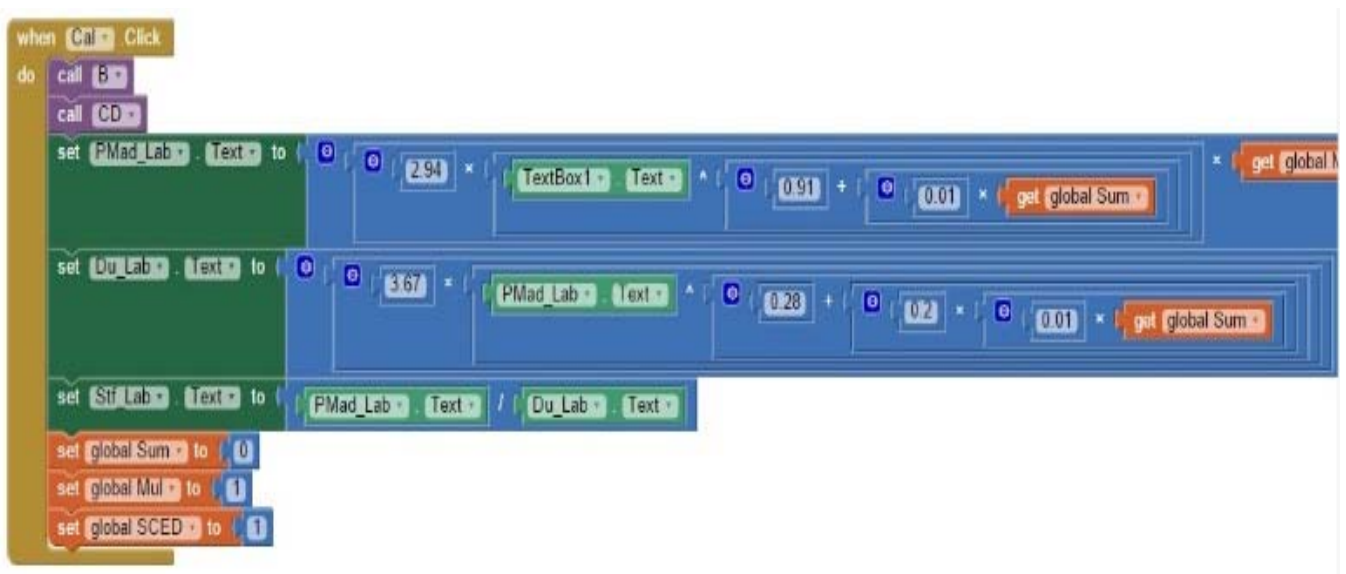

Fig. 7. Code block for CoCoMo II with post architectural model

## B. Results

We executed our Cost Estimation application using android based platform. Visual block programming language is used for the development of application. Fig. 8 to 12 shows the output screens of main application.

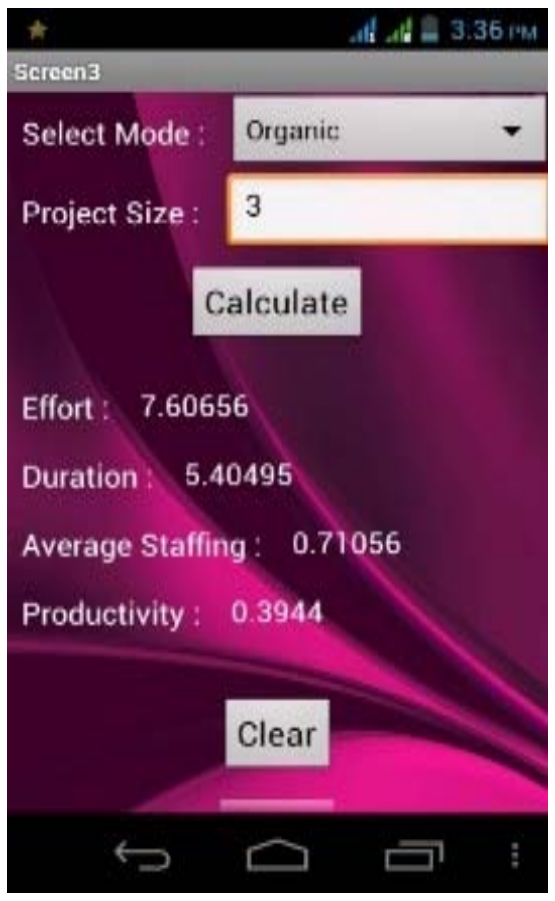

Fig. 8. CoCoMo I with basic model

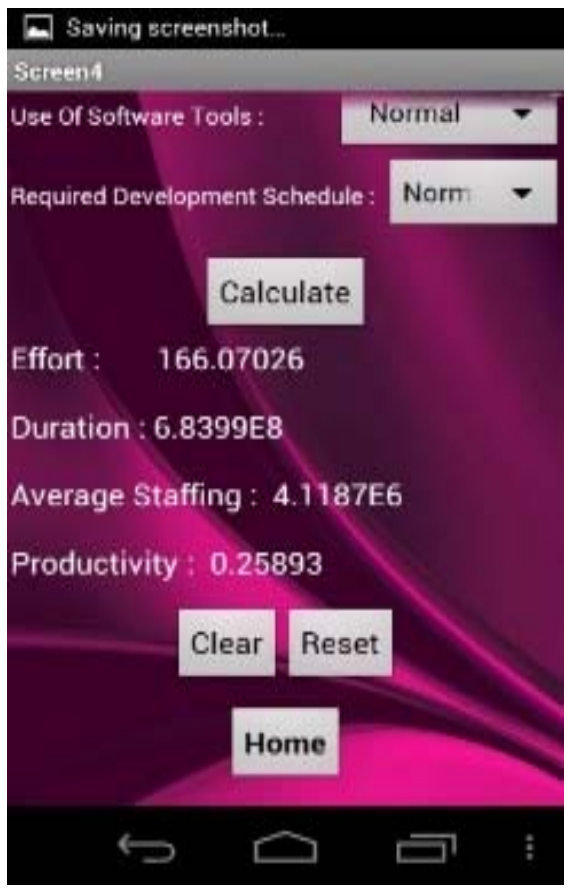

Fig. 9. CoCoMo I with Intermediate model

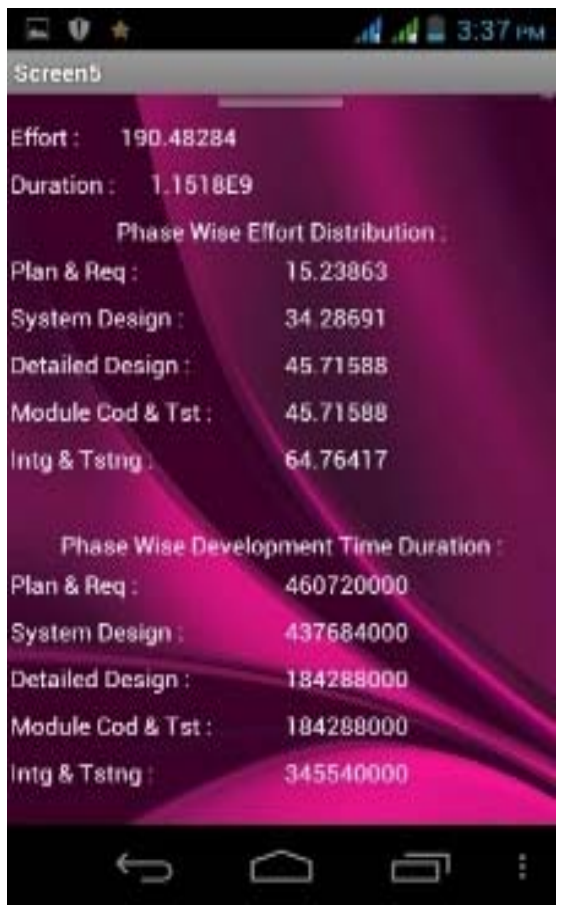

Fig. 10. CoCoMo I with detailed model

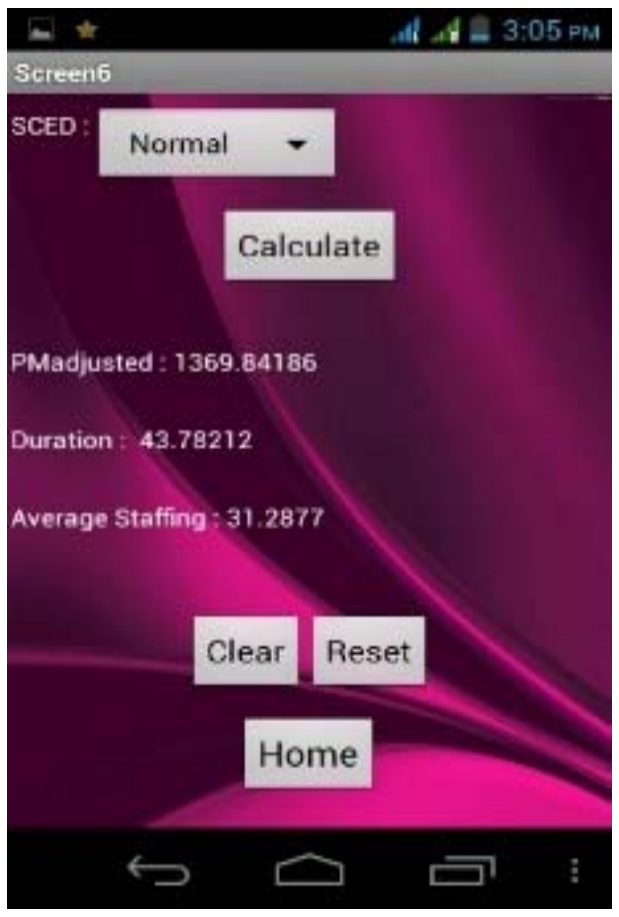

Fig. 11. CoCoMo II with early design model

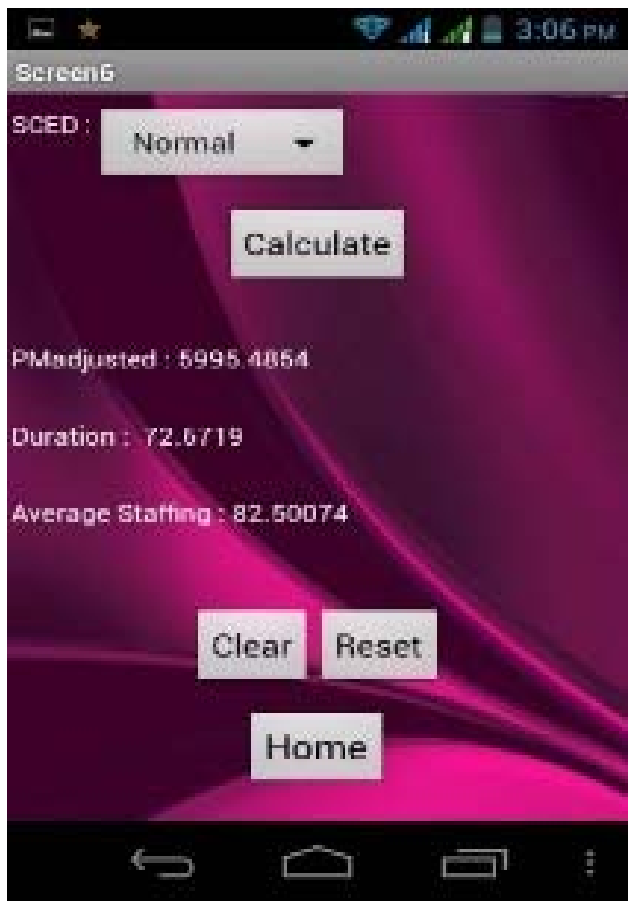

Fig. 12. CoCoMo II with post architecture model

### C. Descriptive Analysis of data

TABLE IV. SHOWING BASIC STATISTICS OF ACCURATE SOFTWARE COST ESTIMATION

| Sr.no | Basic Statistics | | | | |
|---|---|---|---|---|---|
| | *Minimum* | *Maximum* | *Median* | *Mean* | *Standard deviation* |
| 1. | 1.00 | 3.00 | 2.00 | 1.90 | 0.70 |

The minimum and maximum means the smallest and largest number answer choice that collects not less than one response. It is useful to find the range of answer by subtracting the minimum and maximum. In the Table.IV, minimum and maximum of 1 and 3 presents that there were 6 responses in the uppermost answer (i.e. Strongly agreed) and 4 responses in the lowermost answer (i.e. not agreed).The answer choice that is in the center of all responses shows a median, means there are 50% response before median are smaller and 50% response after median are larger. The median of 2.00 (higher than the 1.90 mean) show that there were more respondents who were agreed than respondents who were strongly agreed. The mean gives the average of entire responses by adding all number answer choices and then divide them by total amount of number. In this case, a mean of 1.90 represents the overall respondents came in somewhere between strongly agreed and the agreed. Finally, the standard deviation shows the growth or alteration of your responses, so here the standard deviation is 0.70.

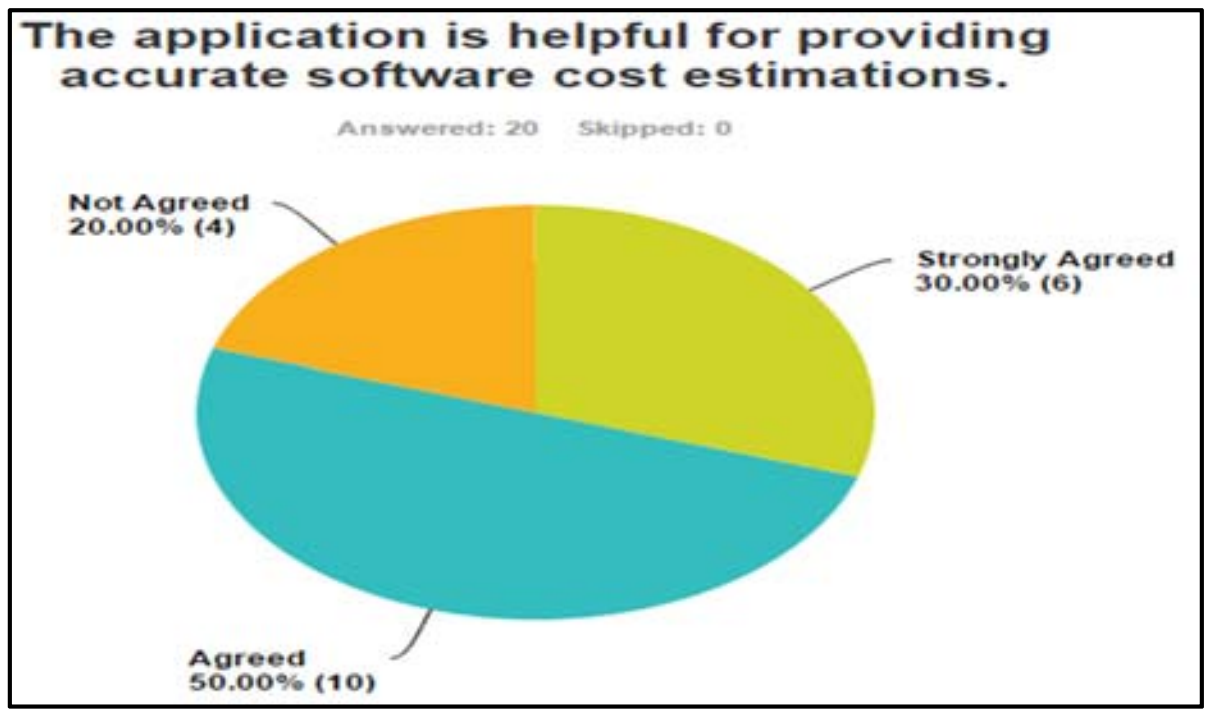

Fig. 13. Pie chart of accurate software cost estimation

Fig. 13 shows that 30% respondents were strongly agreed with the statement, the respondents who were just agreed with the statement were 50% and 20% were not agreed with statement.

Prevent from significant loss: The objective of this question was to get information from the respondent that the application is helpful to prevent them know about the significant loss.

TABLE V. SHOWING BASIC STATISTICS OF THE PROJECT EFFORT

| Sr.no | Basic Statistics | | | | |
|---|---|---|---|---|---|
| | *Minimum* | *Maximum* | *Median* | *Mean* | *Standard deviation* |
| 1. | 1.00 | 3.00 | 2.00 | 1.80 | 0.60 |

In theTable.V, minimum and maximum of 1 and 3 presents that there were 6 responses in the uppermost answer (i.e. Strongly agreed) and 2 responses in the lowermost answer (i.e. not agreed).The median of 2.00 (higher than the 1.80 mean) show that there were more respondents who were agreed than respondents who were strongly agreed. The mean gives the average of all responses. In this case, a mean of 1.80 represents the overall respondents came in somewhere between strongly agreed and the agreed. Finally, the standard deviation shows the growth or alteration of your responses, so here the standard deviation is 0.60.

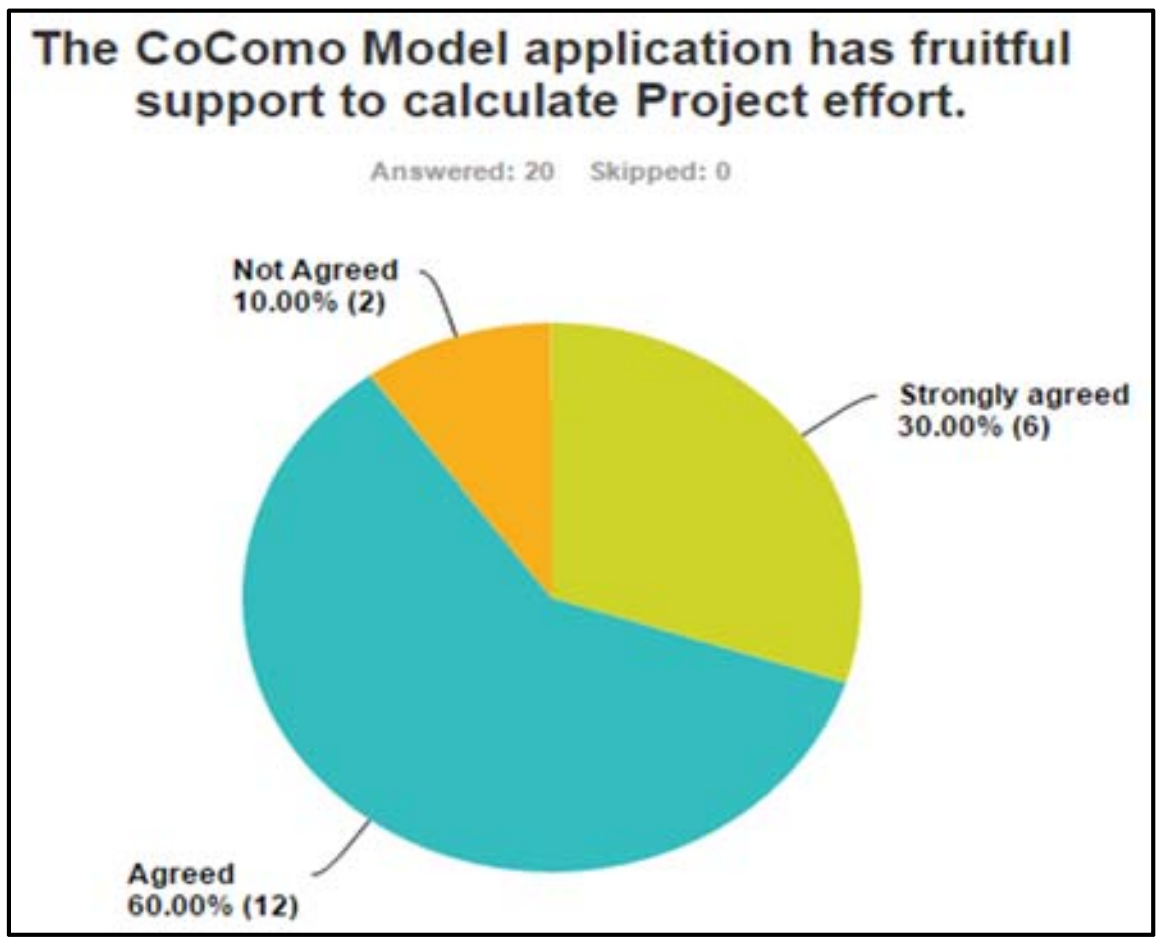

Fig. 14. Pie chart of project effort

The above Fig. 14 shows that the respondents that were strongly agreed with the statement were 30%, whereas 60% respondents were agreed that the application provide support to calculate project effort, while among the 20 respondents feedback 10% respondents were not agreed with the statement.

TABLE VI. SHOWING BAISC STATISTICS OF PREVENT LOSS

| Sr.no | Basic Statistics | | | | |
|---|---|---|---|---|---|
| | *Minimum* | *Maximum* | *Median* | *Mean* | *Standard deviation* |
| 1. | 1.00 | 3.00 | 2.00 | 1.80 | 0.60 |

In the Table.VI, minimum and maximum of 1 and 3 presents that there were 5 responses in the uppermost answer(i.e.

Strongly agreed ) and 4 responses in the lowermost answer(i.e. not agreed).The median of 2.00 (higher than the 1.95 mean) show that there were more respondents who were agreed than respondents who were strongly agreed. In this case, a mean of 1.95 represents the overall respondents came in somewhere between strongly agreed and the agreed. Finally, the standard deviation shows the growth or alteration of your responses, so here the standard deviation is 0.60.

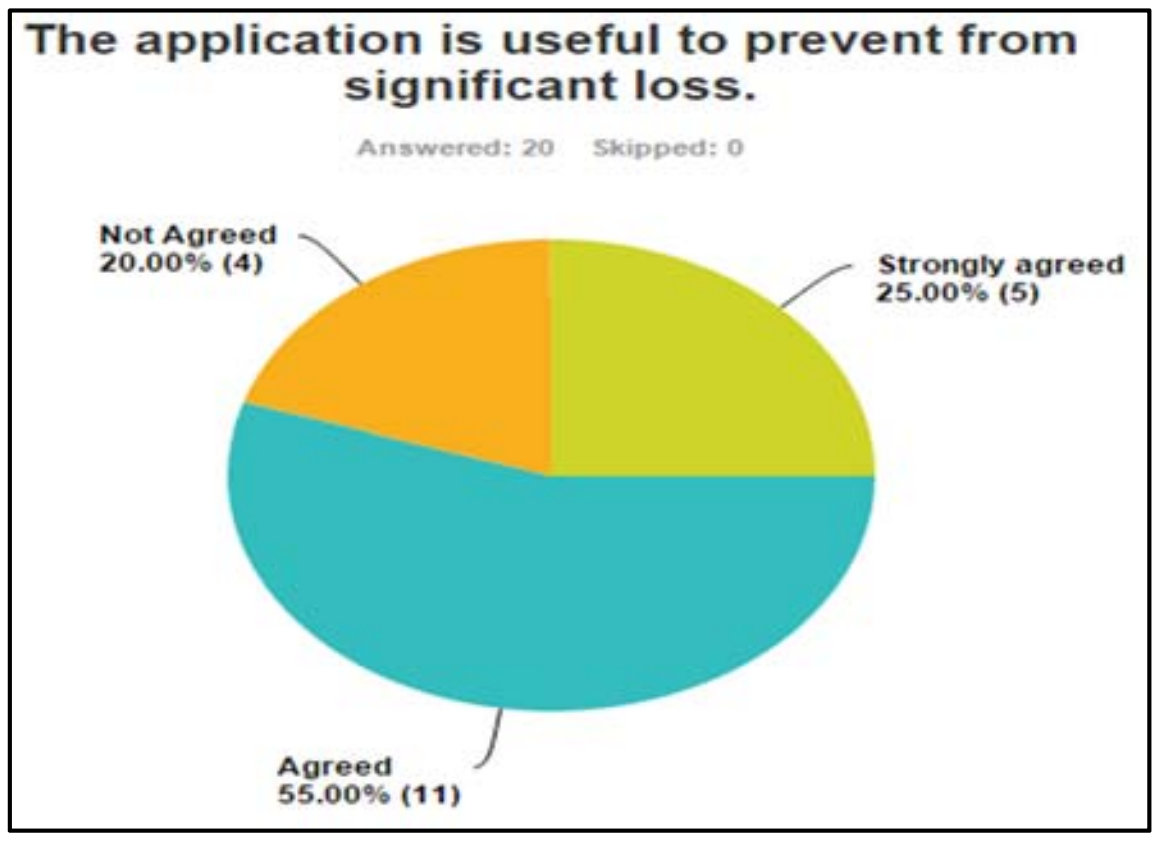

Fig. 15. Pie chart of prevent from significant loss

The above Fig. 15 shows that the respondents that were strongly agreed with the statement were 25%, Agreed respondents were 55% whereas 20 % respondents were not agreed.

TABLE VII. SHOWING BAISC STATISTICS OF THE FINDING APPROXIMATE SLOUTION

| Sr.no | Basic Statistics | | | | |
|---|---|---|---|---|---|
| | ***Minimum*** | ***Maximum*** | ***Median*** | ***Mean*** | ***Standard deviation*** |
| 1. | 1.00 | 3.00 | 2.00 | 2.00 | 0.65 |

In theTable.VII, minimum and maximum of 1 and 3 presents that there were 4 responses in the uppermost answer (i.e. Strongly agreed) and 4 responses in the lowermost answer (i.e. not agreed). The median of 2.00 (equal to the 2.00 mean) show that there were equal number of respondents who were agreed with the statement. In this case, a mean of 2.00 represents the overall respondents came in agreed. Finally, the standard deviation shows the growth or alteration of your responses, so here the standard deviation is 0.65.

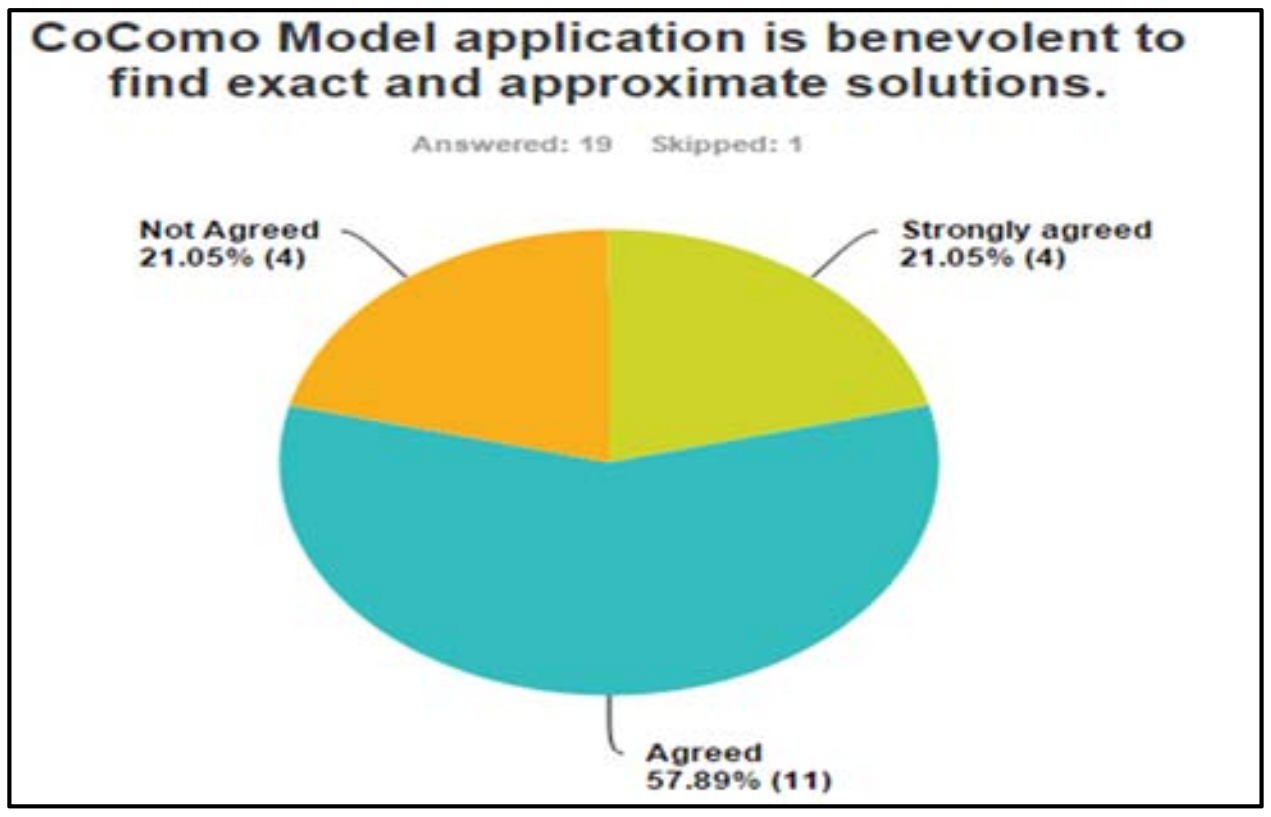

Fig. 16. Pie chart of finding approximate solutions

Fig. 16 shows that the 21.05% respondents were strongly agreed that application is beneficial in terms of finding the exact solutions, 57.89 % respondents were agreed with the statement and 21.05% respondents were not agreed that the application is helpful to find approximate results.

TABLE VIII. SHOWING BASIC STATISTICS OF THE FRIENDLY UEER INTEFACE

| Sr.no | Basic Statistics | | | | |
|---|---|---|---|---|---|
| | ***Minimum*** | ***Maximum*** | ***Median*** | ***Mean*** | ***Standard deviation*** |
| 1. | 1.00 | 3.00 | 2.00 | 1.70 | 0.56 |

In theTable.VIII, minimum and maximum of 1 and 3 presents that there were 7 responses in the uppermost answer (i.e. Strongly agreed) and 1 responses in the lowermost answer (i.e. not agreed). The median of 2.00 (higher than the 1.70 mean) show that there were more respondents who were agreed than respondents who were strongly agreed. A mean of 1.70 represents the overall respondents came in somewhere between strongly agreed and the agreed. Finally, the standard deviation shows the growth or alteration of your responses, so here the standard deviation is 0.56.

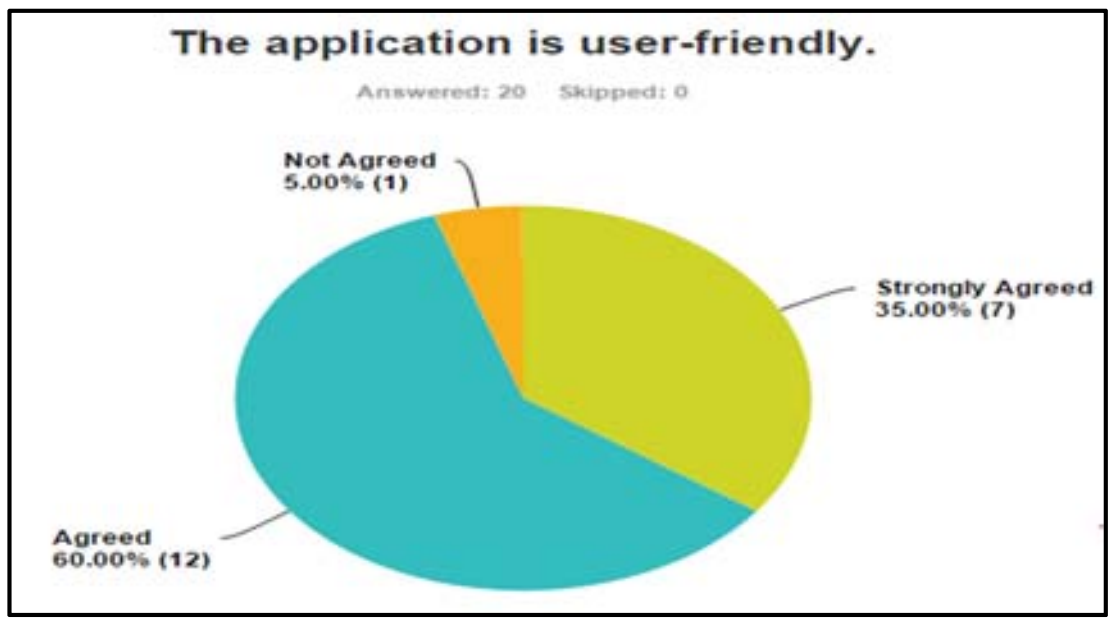

Fig. 17. Pie chart of friendly user interface

Fig. 17 shows that respondents that likes the user interface of the application were 35%, however mostly 60% respondents were found the application is user friendly and only 5% respondents were disagree with the statement.

TABLE IX. SHOWING BASIC STATISTICS OF THE APPROPRIATE EFFORT DURATION

| Sr.no | Basic Statistics | | | | |
|---|---|---|---|---|---|
| | ***Minimum*** | ***Maximum*** | ***Median*** | ***Mean*** | ***Standard deviation*** |
| 1. | 1.00 | 3.00 | 2.00 | 1.90 | 0.54 |

In theTable.IX, minimum and maximum of 1 and 3 presents that there were 4 responses in the uppermost answer(i.e. Strongly agreed ) and 2 responses in the lowermost answer(i.e. not agreed). The median of 2.00 (higher than the 1.90 mean) show that there were more respondents who were agreed than respondents who were strongly agreed. A mean of 1.90 represents the overall respondents came in somewhere between strongly agreed and the agreed. Finally, the standard deviation shows the growth or alteration of your responses, so here the standard deviation is 0.54

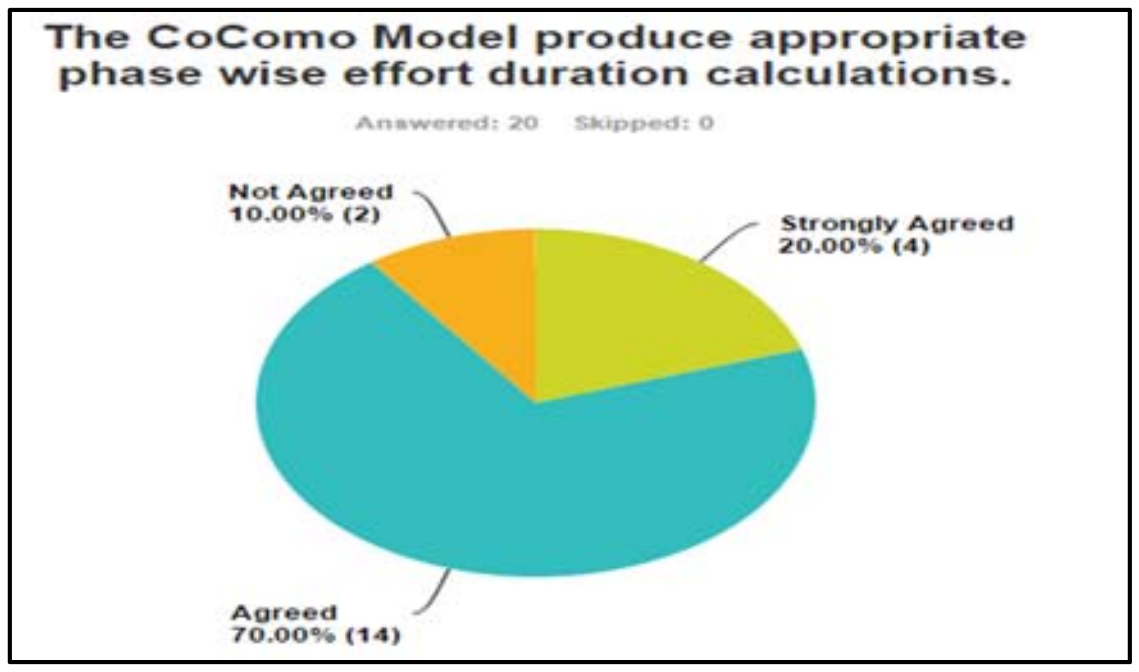

Fig. 18. Pie chart of appropriate effort duration

In the above Fig. 18 shows that 20% respondents were strongly agreed with the statement, the respondents that were agreed with question were 70% and there were just 10% respondents that disagrees with the statement

TABLE X. SHOWING BASIC STATISTICS OF THE UNDERSTANDING OF COCOMO MODEL

| Sr.no | Basic Statistics | | | | |
|---|---|---|---|---|---|
| | ***Minimum*** | ***Maximum*** | ***Median*** | ***Mean*** | ***Standard deviation*** |
| 1. | 1.00 | 3.00 | 2.00 | 1.95 | 0.59 |

In theTable.X, minimum and maximum of 1 and 3 presents that there were 4 responses in the uppermost answer (i.e. Strongly agreed) and 3 responses in the lowermost answer (i.e. not agreed). The median of 2.00 (higher than the 1.95 mean) show that there were more respondents who were agreed than respondents who were strongly agreed. In this case, a mean of 1.95 represents the overall respondents came in somewhere between strongly agreed and the agreed. Finally, the standard deviation shows the growth or alteration of your responses**,** so here the standard deviation is 0.59

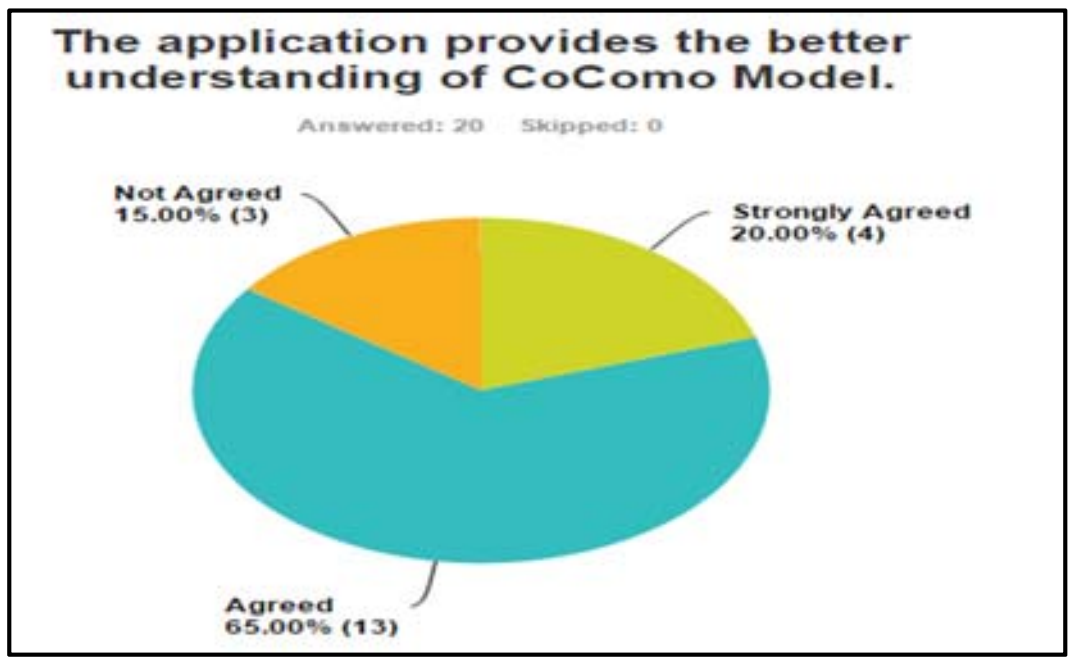

Fig. 19. Pie chart of understanding of CoCoMo model

Fig. 19 shows that strongly agreed respondent with the statement were 20%, 65% respondents were agreed to the statement, while only 15% respondent do not agree with the statement.

TABLE XI. SHOWING BASIC STATISTICS OF DEVELOPMENT MODE

| Sr.no | Basic Statistics | | | | |
|---|---|---|---|---|---|
| | ***Minimum*** | ***Maximum*** | ***Median*** | ***Mean*** | ***Standard deviation*** |
| 1. | 1.00 | 3.00 | 2.00 | 1.94 | 0.64 |

In the Table. XI, minimum and maximum of 1and 3 presents that there were 4 responses in the uppermost answer (i.e. strongly agreed) and 3 responses in the lowermost answer (i.e. not agreed). The median of 2.00 (higher than the 1.94 mean) show that there were more respondents who were agreed than respondents who were strongly agreed. In this case, a mean of 1.94 represents the overall respondents came in somewhere between strongly agreed and the agreed. Finally, the standard deviation shows the growth or alteration of your responses**, so** here the standard deviation is 0.64.

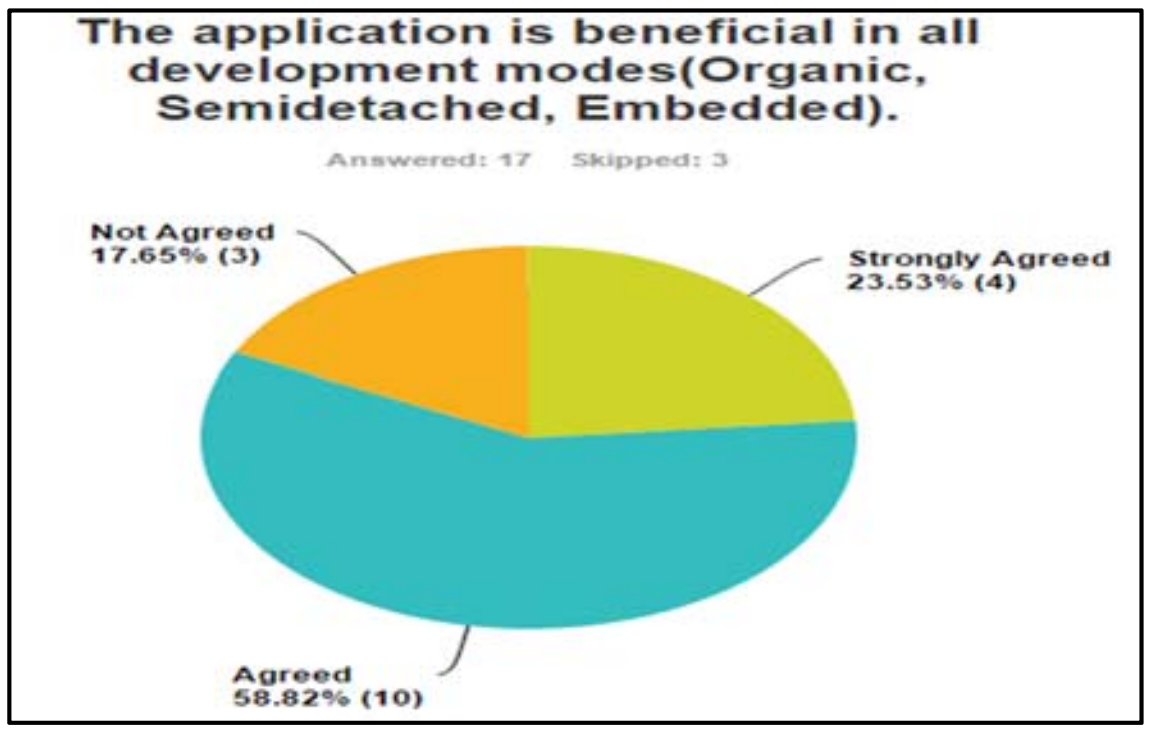

Fig. 20. Pie Chart of development modes

Fig. 20 shows that the respondents that were strongly agreed with the statement were 25.53%, however 58.82% respondents were agreed that the application is benevolent in all development modes and 17.65% respondents were not agreed to the statement

TABLE XII. SHOWING BASIC STATISTICS OF COCOMO I AND COCOMO II

| Sr.no | Basic Statistics | | | | |
|---|---|---|---|---|---|
| | ***Minimum*** | ***Maximum*** | ***Median*** | ***Mean*** | ***Standard deviation*** |
| 1. | 1.00 | 3.00 | 1.00 | 1.60 | 0.73 |

In theTable.XII, minimum and maximum of 1 and 3 presents that there were 11 responses in the uppermost answer (i.e. Strongly agreed) and 3 responses in the lowermost answer (i.e. not agreed). The median of 2.00 (higher than the 1.60 mean) show that there were more respondents who were agreed than respondents who were strongly agreed. In this case, a mean of 1.60 represents the overall respondents came in somewhere between strongly agreed and the agreed. Finally, the standard deviation shows the growth or alteration of your responses, so here the standard deviation is 0.73.

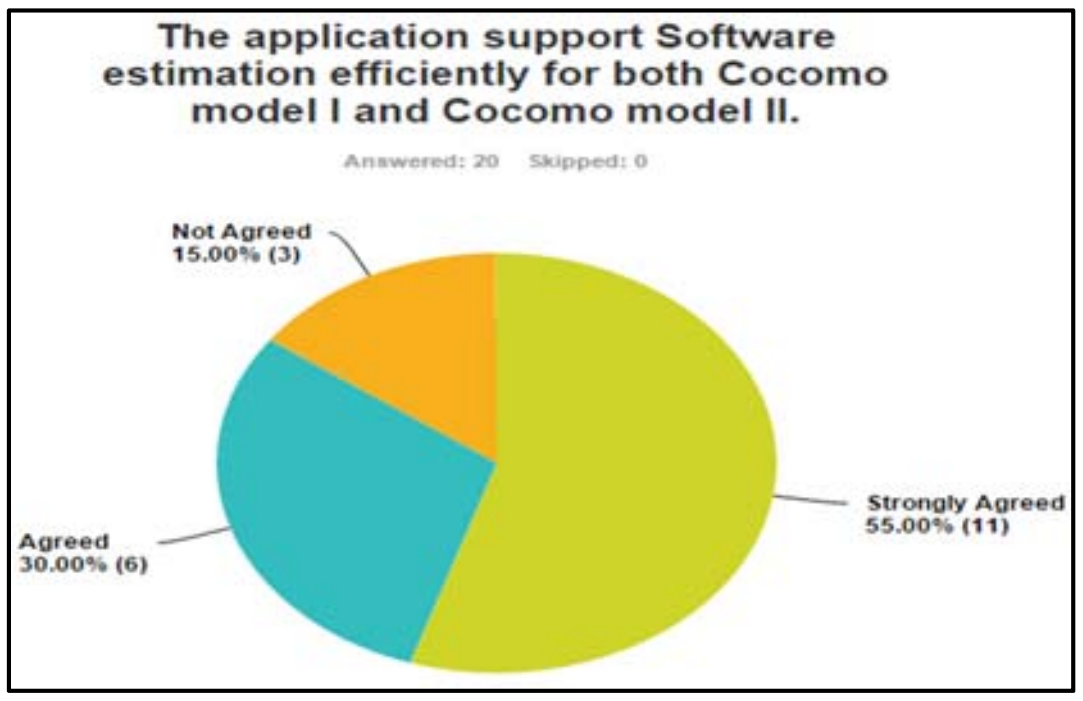

Fig. 21. Pie chart of CoCoMo I and CoCoMo II

In Fig. 21 shows that 55% responses were gathered that strongly agreed with the statement, whereas 30% Reponses from the respondents were agreed with the statement, and only 15% respondents disagrees with the statement.

TABLE XIII. SHOWING BASIC STATISTICS OF THE RATE APPLICATION

| Sr.no | Basic Statistics | | | | |
|---|---|---|---|---|---|
| | *Minimum* | *Maximum* | *Median* | *Mean* | *Standard deviation* |
| 1. | 1.00 | 3.00 | 2.00 | 2.00 | 0.65 |

In the Table.XIII, minimum and maximum of 1 and 3 presents that there were 4 responses in the uppermost answer (i.e. High quality) and 4 responses in the lowermost answer (i.e. Low quality). The median of 2.00 (equal to the 2.00 mean) show that there were equal number of respondents who were agreed with the statement. In this case, a mean of 2.00 shows that overall respondents came in good quality. Finally, the standard deviation shows the growth or alteration of your responses**,** so Here the standard deviation is 0.65.

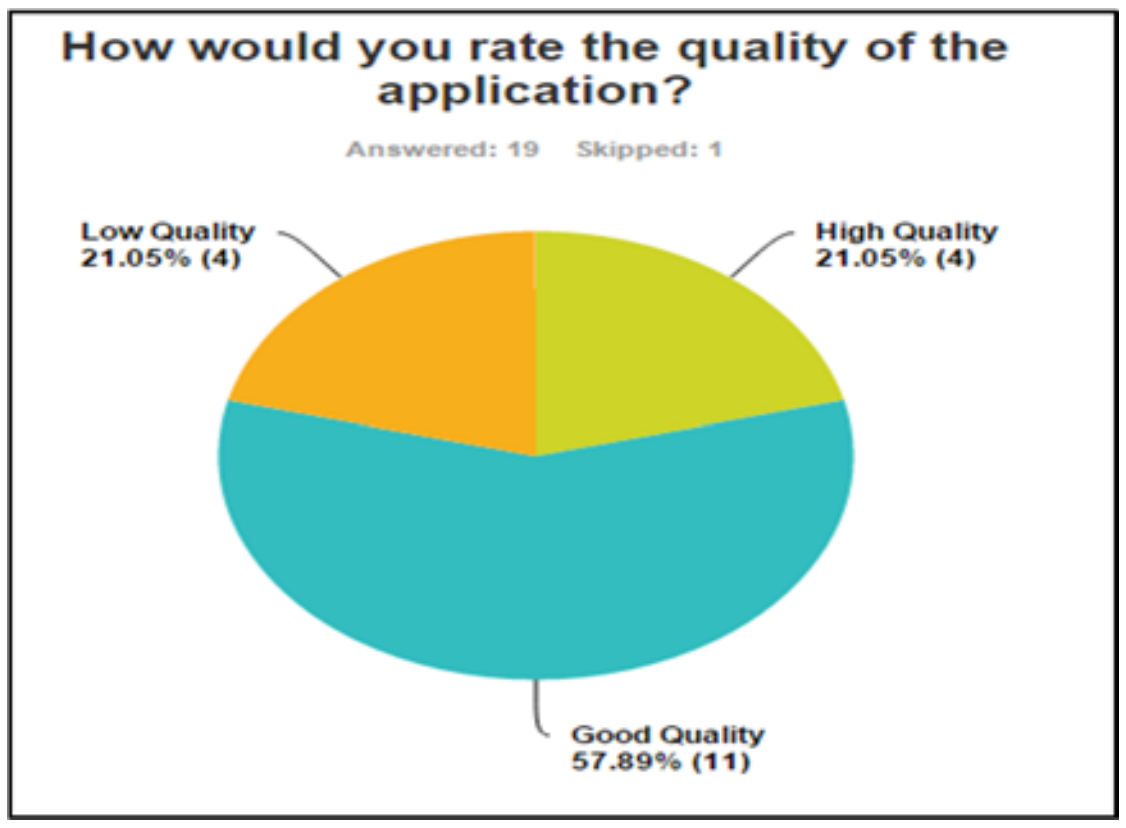

Fig. 22. Pie chart of rate application

Fig. 22 shows respondent that said that the application is of high quality were 21.05%, while 57.89% were said that the application is of high quality and 21.05% respondent found that the application is of low quality.

## IV. CONCLUSIONS

This study deals with development of android-based estimation application with focus on developing and integrating the module, name estimation.

The estimation module of our application is quite important with respect to facilitation the students of software engineering for performing CoCoMo-based cost estimation. The application covers both versions of CoCoMo, namely (1) CoCoMo-1, and (2) CoCoMo-II. The distinctive feature of this module is that, to the best of our knowledge, no such application exists in the android-based paradigm. The developed models provide estimation results.

The proposed application is efficient with respect to estimation by using block-oriented programming technique.

***Future directions:*** The application can be enhanced as follows: To implement remaining two models of Cocomo 2, namely (1) application composition model, (2) reuse model.

## ACKNOWLEDGMENT

In the name of Allah, the Most Merciful and the Most Gracious, Alhamdulillah, all praises to Allah for the strengths and His Blessings in completing this research paper.

We would like to express our deepest gratitude to our Supervisor Dr. Zubair for his excellent guidance, patience, caring and providing us with an excellent atmosphere for doing our research work